\documentclass[12pt,a4paper]{article}

\usepackage{amsfonts}

\newcommand{\Ro}{{\mathbb R}}

\newcommand{\be}{\begin{eqnarray}}
\newcommand{\ee}{\end{eqnarray}}

\begin{document}

\title {Stability of families of probability distributions
under reduction of the number of degrees of freedom}

\author{Christophe Vignat$^*$ and Jan Naudts$^\dagger$\\
{\small $^*$ Equipe Signal et Communications,
Universit\'e de Marne la Vall\'ee,}\\
{\small 77454 Marne la Vall\'ee Cedex 2}\\
{\small $^\dagger$ Departement Fysica, Universiteit Antwerpen, Belgium}\\
{\small E-mail: vignat@univ-mlv.fr, jan.naudts@ua.ac.be}
}

\date {April 2004}
\maketitle

\begin{abstract}
We consider two classes of probability distributions for configurations of the ideal gas.
They depend only on kinetic energy and they remain of the same form
when degrees of freedom are integrated out. The relation
with equilibrium distributions of Tsallis' thermostatistics is discussed.
\end{abstract}

\section{Introduction}

This paper studies two classes of probability distributions for the
ideal gas with $n$ degrees of freedom. Their main property is that
elimination of some of the degrees of freedom produces again a
distribution of the same class.

Particle velocities in an ideal gas with infinitely many degrees of freedom,
in thermodynamic equilibrium at inverse temperature $\beta$,
follow the Maxwell distribution. This result is
known for more than a century. But quite often one is interested in
finite systems or in systems out of equilibrium. For such systems
deviations from a Gaussian distribution are expected.
Such deviations have been reported recently \cite {JESK03}
in laser-cooled dilute samples of atomic gases.
The velocity distribution of migrating Hydra cells in two-dimensional aggregates
is non-Gaussian, the mean-square displacement is superdiffusive \cite {URGS01}.
Numerical simulation of metastable states in a long-ranged mean-field model
\cite {LR01,LRT01,LRT02} reveal non-Gaussian velocity distributions.
Many more examples should be known but a systematic investigation
seems to be missing.

If the velocity distribution is not Maxwellian then one cannot expect
it to be of the product form. Indeed, the standard argument
\be
\exp\bigg(-(\beta/2)\sum_jm|v_j|^2\bigg)
&=&\prod_j \exp(-(\beta/2)m|v_j|^2)
\ee
works only for the exponential function. This raises the question
whether one can formulate alternatives which are still easy to manage.
Distributions (\ref {pne}, \ref {qne}), studied in the present work,
satisfy the requirement of simplicity. They depend only on the
total energy of the gas. They have the additional property that,
if some of the degrees of freedom are integrated out, then the
remaining degrees of freedom obey a probability distribution
of the same form. Two classes of distributions are introduced.
They correspond with the cases $q>1$ and $q<1$ in the terminology of
non-extensive thermostatistics. The first class contains distributions
for which the probability of large energies decays algebraically
instead of exponentially. The other class consists of distributions
with a cutoff for large energies.

In the next section the two classes of probability distributions
are derived starting from simple models for the velocity variables.
In Section 3 is shown that the distributions depend only on total
energy and a duality relation between the two classes is pointed out.
In Section 4 the relation with equilibrium distributions of non-extensive
thermostatistics is explained. The latter do not generically satisfy
the requirement of stability under reduction of degrees of freedom.
In the final section follows a discussion of our results.

\section{Two families of probability distributions}

A characteristic feature of the ideal gas with $n$ degrees
of freedom is that it is described by independent variables
$k_1,k_2,\cdots,k_n$, components of the velocities
of the particles.
It is tradition to assume that each of the variables $k_j$ obeys
the Maxwell distribution
\be
\sqrt{\frac {\beta m_j}{2\pi}}\,
\exp\left(-\frac 12\beta m_jk_j^2\right)
\label {m}
\ee
with $\beta$ the inverse temperature and with $m_j$ the mass of the $j$-th degree
of freedom.  A slight generalization is obtained by replacing the Gaussian
distribution $(\lambda/\pi)^{1/2}\exp(-\lambda u^2)$ by an exponential power distribution
\be
p^{\rm EP}_{z,\lambda}(u)=\frac {z \lambda^{1/z}}{2\Gamma(1/z)}\exp(-\lambda |u|^z),
\label {pe}
\ee
with $\lambda$ the scale parameter and $z$ the shape parameter.
These distributions were introduced by Subbotin \cite {SMT23} in 1923.
The distribution (\ref {m}) is then replaced by $p^{\rm EP}_{z,\lambda_j}(k_j)$
with
\be
\lambda_j=\frac 12\beta m_j.
\ee
This generalization is not essential for the paper. One can take $z=2$ in
all what follows.

Several arguments can be invoked to explain why sometimes the distribution of
experimentally observed velocities $v_j$ differs from that of the independent variables $k_j$.
The simplest argument is that the distribution is modulated by an independent variable $s$.
The relation between $v$ and $k$ is assumed to be of the form
\be
v_j=\frac {g(\beta)^{1/z}k_j}{s^{1/z}}
\label {vks}
\ee
with $g(\beta)$ a positive proportionality constant, depending on inverse temperature $\beta$.
We will assume that $s$ obeys a Gamma-distribution
with shape parameter $b+1$
\be
w(s)=\frac 1{\Gamma(b+1)}s^be^{-s}.
\ee
The probability distribution of the velocities $v$ is then found to be
\be
 p_n(v)=\int_0^{+\infty}{\rm d}s\,w(s)\prod_{j=1}^np^{\rm EP}_{z,s\lambda_j/g(\beta)}(v_j).
\label {pdl}
\ee

An interesting alternative for (\ref {vks}) is
\be
v_j=\frac {g(\beta)^{1/z} k_j}{\left(s_n+\sum_{l=1}^n\lambda_l|k_l|^z\right)^{1/z}}
\label {trvsk}
\ee
where $s_n$ has a Gamma-distribution with shape parameter $b_n$.
These probability distributions form a family stable under reduction
of degrees of freedom because
\be
s_{n-1}=s_n+\lambda_n |k_n|^z
\ee
has again a Gamma-distribution, this time with shape parameter $b_{n-1}$ given by
\be
b_{n-1}=b_n+\frac 1z
\label {bofn}
\ee
For further use let $b_n=b-1-n/z$.
The resulting probability distribution for the velocities $v_j$, given by (\ref {trvsk}), is denoted $q_n(v)$
and, after some calculation, is found to be
\be
q_n(v)
&=&\int_0^{+\infty}{\rm d}s_n\,w_n(s_n)
J_n(s_n,k)
\prod_{j=1}^np_{z,\lambda_j}^{\rm EP}(k_j)
\label {pds}
\ee
with $k$ related to $v$ by (\ref {trvsk}), with
\be
w_n(s)&=&\frac 1{\Gamma(b_n)}s^{b_n-1}e^{-s}
\ee
and with
\be
J_n(s_n,k)=\det\left(\frac {\partial k}{\partial v}\right)
=\left(\frac {g(\beta)^{1/z}}{\left[s_n+\sum_{l=1}^n\lambda_l |k_l|^z\right]^{1/z}}\right)^{-n}.
\ee

\section{Energy distribution and duality}

The energy of the ideal gas of $n$ particles is given by
\be
H_n(v)=\frac 12\sum_{j=1}^nm_j\sum_{\alpha=1}^\nu|v_{j,\alpha}|^{z}.
\ee
The probability distributions (\ref {pdl}) and (\ref {pds}) depend only
on the velocities $v$  via the energy $E=H_n(v)$.
Indeed, a short calculation gives
\be
 p_n(v)\equiv p_n(E)&=&\frac 1{Z_n(\beta)}\frac 1{\left[g(\beta)+\beta E\right]^{b+1+n/z}}
\label {pne}\\
 q_n(v)\equiv q_n(E)&=&\frac 1{\zeta_n(\beta)}\left[g(\beta)-\beta E\right]_+^{b-1-n/z}
\label {qne}
\ee
with
\be
Z_n(\beta)&=&\left(\frac {2\Gamma(1/z)}z\right)^n\frac 1{\prod_{j=1}^n\lambda_j^{1/z}}g(\beta)^{b+1}\cr
\zeta_n(\beta)&=&\frac {\Gamma(b-1-n/z)}{\Gamma(b-1)}\left(\frac {2\Gamma(1/z)}z\right)^n
\frac 1{\prod_{j=1}^n\lambda_j^{1/z}} g(\beta)^{b-1}.
\ee
The notation $[u]_+=\max\{0,u\}$ is used.
Note that $n<z(b-1)$ is required for $q_n(v)$ to be well-defined.

Let us now clarify the relation between $p_n(E)$ and $q_n(E)$. First note that
expression (\ref {pne}) is meaningful for negative $n$, as long as $z(b+1)+n>0$ is satisfied.
Similarly, (\ref {qne}) is meaningfully for all negative $n$. The transformation
\be
E'=\frac {g(\beta)E}{g(\beta)+\beta E}
\ee
maps $p_{-n}(E)$ onto $q_n(E')$. Indeed, one has
\be
p_{-n}(E)&\rightarrow&
\frac {{\rm d}E}{{\rm d}E'}\,p_{-n}(E)\cr
&=&\frac {g(\beta)^2}{(g(\beta)-\beta E')^2}
\frac 1{Z_{-n}(\beta)}\left[g(\beta)+\frac {\beta g(\beta)E'}{g(\beta)-\beta E'}\right]^{-b-1+n/z}\cr
&=&\frac 1{Z_{-n}(\beta)}\frac {(g(\beta)-\beta E')^{b-1-n/z}}{g(\beta)^{2(b+1-n/z)}}\cr
&=&q_n(E')
\ee
provided that the normalization constants $Z_n(\beta)$ and $\zeta_n(\beta)$ for negative values of $n$
are defined in an appropriate way.

\section{Non-extensive thermostatistics}

In the context of Tsallis' non-extensive
thermostatistics various proposals have been made for the equilibrium
probability distribution at inverse temperature $\beta$.
One of these reads \cite {TMP98}
\be
p(E)=\frac 1{Z(\beta)}\left[1-(1-q)\beta\frac {E-U_q(\beta)}{Z(\beta)^{1-q}}
\right]_+^{\frac 1{1-q}}.
\label {tmp}
\ee
with normalization factor $Z(\beta)$ and with
\be
U_q(\beta)=\frac 1{Z(\beta)^{1-q}}\int{\rm d}E\,p(E)^qE.
\ee
In the limit that
the parameter $q$ equals 1 then (\ref {tmp}) reduces to the well-known Boltzmann-Gibbs distribution.
Identification of (\ref {tmp})  with  (\ref {pne}) gives
\be
g(\beta)&=&\frac {Z(\beta)^{q-1}}{q-1}-\beta U_q(\beta)\cr
1+b+\frac nz&=&\frac 1{q-1}.
\ee
Identification with (\ref {qne}) gives
\be
g(\beta)&=&\frac {Z(\beta)^{1-q}}{1-q}+\beta U_q(\beta)\cr
b-1-\frac nz&=&\frac 1{1-q}.
\ee
These relations show that the parameter $q$ must depend on the number of particles $n$.
In general this identification is only possible for a single value of
$n$ because the function $g(\beta)$ should not depend on $n$.

Recently, one of the authors \cite {NJ03,NJ04} has proposed equilibrium distributions of the form
\be
p_n(E)=\exp_\phi(G(\beta)-\beta E)
\ee
with $\exp_\phi(u)$ the inverse function of
\be
\ln_\phi(u)=\int_1^x{\rm d}v\,\frac 1{\phi(v)},
\ee
where $\phi(v)$ is an arbitrary positive function.
Identification with (\ref {pne}, \ref {qne}) is only possible
with a function $\phi$ which depends on $\beta$ and on $n$.

One concludes that in general the probability distributions used
in non-extensive thermostatistics are not compatible with
the reduction requirement which is central to the present work.

\section{Discussion}

Two classes of probability distributions for the velocities of an ideal gas have been introduced.
The distributions $p_n(v)$ given by (\ref {pne}) have the property
that integrating out one of the velocities produces a probability distribution of the
same class
\be
\int_\Ro{\rm d}v_n\,p_n(v)=p_{n-1}(v).
\label {reduc}
\ee
The same property holds for the distributions $q_n(v)$ given by (\ref {qne}).
In addition, both kinds of probability distributions depend only on total kinetic energy.

For large values of energy $E$ the probability distribution $p_n(v)$ behaves
as
\be
\frac 1{E^{b+1+n/z}}
\ee
with constants $b$ and $z$ (usually is $z=2$). Because of the $n$-dependence
of the exponent the distribution of large systems approaches the Maxwell distribution.
On the other hand, the probability distribution $q_n(v)$ vanishes for $E$
larger than the cutoff value $g(\beta)/\beta$. In addition, $q_n(v)$ is only defined
for a number of particles less than some maximum value $z(b-1)$. Hence it
is not possible to consider the thermodynamic limit $n\rightarrow +\infty$.

Our derivation of the probability distribution $p_n(v)$ in Section 2
starts with a model of velocities modulated by a variable which obeys
a Gamma-distribution --- see expression (\ref {vks}). This same model is the basis
of superstatistics \cite {BC01}, which explained distributions
of Tsallis' nonextensive thermostatistics in case $q>1$.
For the case $q<1$ no such explanation was available. Here we
show how a modification of (\ref {vks}) leads to the distribution $q_n(v)$,
which in the Tsallis context corresponds with $q<1$.

In Section 4 is shown that in the generic case the distributions
of Tsallis' nonextensive thermostatistics do not satisfy the
requirement (\ref {reduc}) of stability under reduction.
Note that in the context of finite-size renormalization
a condition weaker than (\ref {reduc}) is considered.
This point of view has been elaborated in \cite {TM01}.


\end{document}